\documentclass[final, 5p, times]{elsarticle}
\DeclareUnicodeCharacter{2113}{\ensuremath{\ell}}
\usepackage{caption}
\usepackage{subcaption}
\usepackage{xcolor}
\usepackage{url} 
\usepackage{amsmath}
\usepackage{enumitem}
\usepackage{graphicx}
\usepackage{geometry}
\journal{arxiv.org}

\begin{document}
\sloppy
\begin{frontmatter}

\title{A GEANT4-Based Simulation of Directional Neutron Detectors Using Liquid Scintillators and Boron Carbide Moderators}

\author[inst1]{J.-H. Chen}
\author[inst1]{M. Mirzakhani}
\author[inst1]{R. Mahapatra}
\author[inst1]{S. Sahoo}


\affiliation[inst1]{organization={Department of Physics and Astronomy, Texas A\&M University },
            addressline={578 University Dr}, 
            city={College Station},
            postcode={77840}, 
            state={TX},
            country={US}}

\begin{abstract}
We present a simulation-based study of a compact directional neutron detector composed of liquid scintillator, Cesium Iodide, with boron carbide (B$_4$C) moderation, and silicon photomultipliers (SiPMs). Using GEANT4, we explored multiple detector geometries and material configurations, finding neutron detection efficiencies ranging from approximately 10\% to 30\%. To evaluate directionality, spatial energy distributions were analyzed and used to train a machine learning classifier, which achieved 100\% accuracy in identifying neutron source directions along four cardinal axes. The model remained effective for sources near detector edges, demonstrating robustness. These results establish the feasibility of the proposed detector for applications in nuclear safety, environmental monitoring, and scientific research applications, with future work focused on experimental validation.
\end{abstract}

\end{frontmatter}

\section{Introduction}

Directional neutron detectors are advanced instruments designed to determine not only the presence of neutrons but also their direction of origin. This capability is crucial for applications in nuclear safety, scientific research, homeland security, and environmental monitoring. By identifying the direction of a neutron source, these detectors enable efficient source localization, imaging, and background discrimination.

Neutrons, being electrically neutral, do not ionize matter directly. Therefore, neutron detection relies on indirect interactions. The most fundamental among these is elastic scattering, particularly off hydrogen or helium nuclei, which produces recoil protons. These recoil protons can be tracked to infer the incoming neutron's direction. Other interactions include inelastic scattering and neutron-gamma (n, $\gamma$) reactions, which emit gamma rays useful for tagging or background rejection. Additionally, neutron-proton (n, p) reactions in some detector materials (e.g., CsI(Tl)) can generate background signals that may mimic true neutron interactions.

To exploit these interactions for directional detection, various detector designs have been developed. One approach uses Time Projection Chambers (TPCs) with scintillation readout to reconstruct three-dimensional recoil proton tracks~\cite{FU2020161445, Jaegle_2019}. Another strategy employs plastic scintillating fiber arrays that record energy deposition patterns of recoil protons~\cite{FU2020161445}. A third class uses recoil proton telescopes, combining hydrogenous radiators, collimators, and scintillation detectors to constrain the incoming neutron direction~\cite{KANEKO1997157}.

Broadly, directional neutron detectors fall into two categories. The first mimics gamma-ray Compton cameras by using multiple scintillators to measure the sequence of neutron interactions and infer directionality~\cite{4178953, brennan2010results}. The second utilizes TPCs to directly image recoil tracks from single elastic scattering events~\cite{fu2020directional}.

This work explores the simulation of several directional neutron detector designs that combine different scintillating materials and silicon photomultipliers (SiPMs) for light readout. Among the configurations studied, one approach uses liquid scintillator (LS) along with a boron carbide ($B_4C$) moderator. In this setup, fast neutrons are first thermalized in the $B_4C$, then captured to produce secondary particles that subsequently interact with the scintillator. This mechanism allows for neutron detection in a compact volume. 

To evaluate the performance and directional sensitivity across different material choices, two alternative configurations are also studied: $B_4C$ combined with CsI, which benefits from CsI's high gamma sensitivity, and an LS-only design without a moderator, which reduces weight and complexity. These three setups, (1) LS+$B_4C$, (2) CsI+$B_4C$, and (3) LS only, are simulated to evaluate their respective detection capabilities and directional sensitivity.

In addition, replacing conventional photomultiplier tubes (PMTs) with SiPMs addresses several practical limitations. SiPMs are significantly smaller, consume less power, and operate at low voltages (typically tens of volts), as opposed to the high-voltage (kilovolt-level) requirements of PMTs. These improvements enable a compact, portable, and robust detector architecture.

As a result of using SiPMs, such a compact and lightweight system has numerous practical advantages. It can be deployed in constrained spaces, mounted on mobile platforms, or even attached to drones for aerial neutron mapping. The mobility of the detector further enhances its capability to localize neutron sources in the field by physically approaching the emission point.

This study primarily focuses on the simulation of the proposed detector design. Using the GEANT4 toolkit, we model the detector geometry and simulate neutron interactions to evaluate energy deposition patterns and spatial distributions. Furthermore, we explore the use of machine learning methods to correlate spatial interaction features with source location, aiming to improve directional resolution. The insights gained from this simulation will guide future design optimizations and prototype development.

\section{Theoretical Framework}
\subsection{Neutron Interaction with Boron Carbide}
\begin{figure}[h!]
  \centering
    \includegraphics[width=\linewidth]{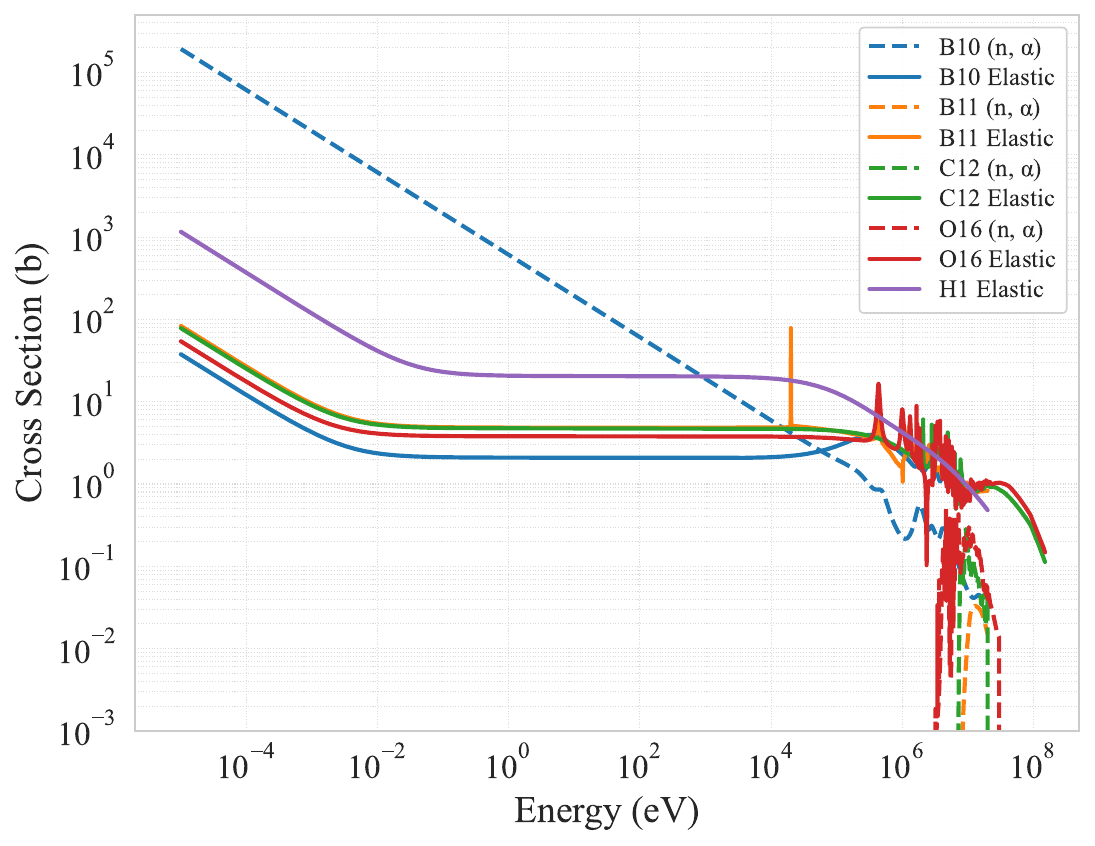}
    \vspace{1em} 
    \includegraphics[width=\linewidth]{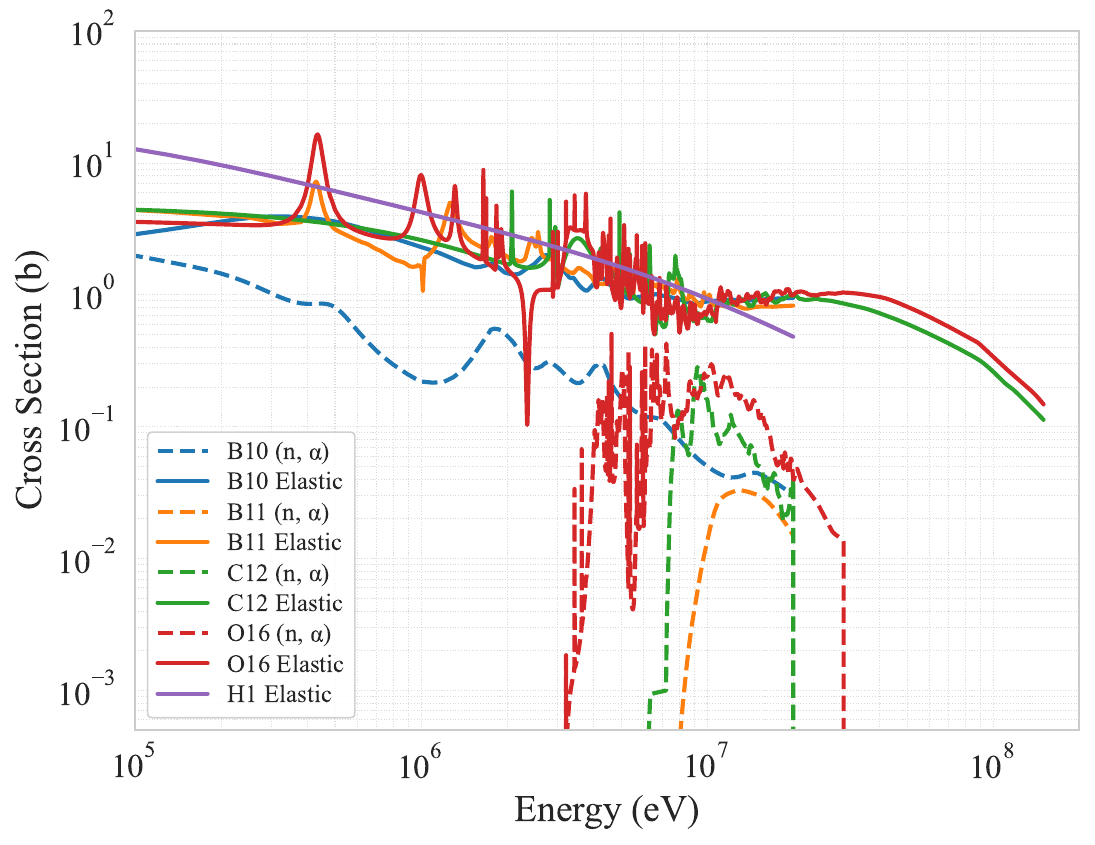}

  \caption{Neutron interaction cross sections from ENDF/B-VIII.0 nuclear data library \cite{brown2018endf} at different energies for selected nuclei (top) and Zoomed-in view, focusing on the cross section at the MeV scale (bottom).  The (n, $\alpha$) reaction and elastic scattering channels are shown for \(\mathrm{^{10}B}\), \(\mathrm{^{11}B}\), \(\mathrm{^{12}C}\), \(\mathrm{^{16}O}\), and \(\mathrm{^{1}H}\).}
  \label{fig:neutron-cross-section-combined}
\end{figure}

The detector design includes a moderator to slow down and absorb incident neutrons. We choose boron carbide as the moderator since it is well known for its high neutron absorption cross section, and is widely used as an absorber in nuclear reactors. In particular, \(\mathrm{^{10}B}\) is quite efficient at absorbing thermal neutrons and fast neutrons \cite{simeone2000study}. Fig.~\ref{fig:neutron-cross-section-combined} show the cross sections of neutron elastic scattering and (n, $\alpha$) reactions with various light elements \cite{soppera2014janis}.

The main interaction channels between neutrons and \(\mathrm{^{10}B}\) are elastic scattering and the boron to lithium (\(\mathrm{^{10}B(n,\:\alpha)^{7}Li}\)) reaction. The (n, $\alpha$) reaction produces an alpha particle and a lithium ion with average kinetic energies of 1.48 and 0.83 MeV, respectively \cite{simeone2000study}, and can be written as follows \cite{wang2022boron, soloway1997rationale}:

\begin{equation} \label{eq: reaction1}
    ^{10}B \: + \: n \: \to \: ^{7}Li \: + \: \alpha \: + \: 2.79\:MeV \quad (6\%) \\
\end{equation}
\begin{equation} \label{eq: reaction2}
    ^{10}B \: + \: n \: \to \: ^{7}Li^* + \: \alpha \: + \: 2.31\:MeV \quad (94\%)
\end{equation}

 Where reaction \ref{eq: reaction2} produces lithium in its first excited state, followed by the emission of a 478 keV gamma ray. The alpha particles and \(\mathrm{^{7}Li}\) ions deposit their kinetic energy locally due to their short range in boron carbide. However, the emitted gamma can travel in longer distance due to the low atomic number of boron carbide.

The 478 keV gamma is a significant source of scintillation light, particularly in setups using CsI, and also contributes to scintillation photon creation in liquid scintillators. However, for MeV-scale neutrons, the \(\mathrm{^{10}B(n,\:\alpha)^{7}Li}\) reaction has a cross section much lower than elastic scattering, as shown in Fig.~\ref{fig:neutron-cross-section-combined}. Therefore, we rely on a series of elastic scatterings to slow down the neutrons, then they can eventually be captured via the (n, $\alpha$) reaction. 

\subsection{Gamma and Neutron Interaction with Liquid Scintillator}

We use liquid scintillator as our active detector material. Liquid scintillators offer several advantages: they are lightweight, their versatility allows for easy customization in shape and size to accommodate various detector geometries, and they have a fast time response, about a few nanoseconds, allowing for fast signal processing. Moreover, they are generally inexpensive and easy to prepare. 

In this detector set-up, the main channels for particle detection are gamma and neutron interactions with the liquid scintillator.  As we know for gamma interactions, particles primarily lose energy through three processes: photoelectric effect, Compton scattering, and pair production. The fraction of incoming gammas that interact through one of these processes as they pass through a scintillator of thickness \(\mathrm{d}\) can be written as:

\begin{equation}
    f \: = \: 1 \: - \: exp(-\mu d)
\end{equation}

 Where $\mu$ is the linear attenuation coefficient in cm$^{-1}$. Specifically, $\mu$ is the total of three linear attenuation coefficients for each interaction, thus:

\begin{equation}
    \mu \: = \: \mu_{photo} \: + \: \mu_{compton} \: + \: \mu_{pair}
\end{equation}

The photoelectric effect is dominant for low energy gammas, usually below 100 keV, but its contribution rapidly decreases as gamma energy increases. Compton scattering dominates at intermediate energies from 100 keV to about 10 MeV. Pair production becomes dominant at higher energies, but its threshold is 1.022 MeV, and it is not relevant for the gamma energies at 478 keV produced by \(\mathrm{^{7}Li}\) de-excitation \cite{fischer2005neutron}. 

Neutrons can undergo elastic scattering with all nuclei present in the detector materials. The recoil energy \( \mathrm{E_r} \) transferred to the target nucleus is given by:

\begin{equation}
    E_{r} \: = \: \frac{4A}{(1+A)^{2}}E_{n}cos^{2}\theta
\end{equation}

Where \( \mathrm{E_n} \) is the energy of the incident neutron, $A$ is the mass number of the target nucleus and $\theta$ is the recoil angle in lab frame \cite{1965Birks}. For a head-on collision ($\theta$ = 0), it is clear hitting a hydrogen atom, i.e. the (n, p) process, will result in the maximum energy transfer, \( \mathrm{E_r = E_n} \) when $A$ = 1. For carbon and oxygen nuclei in the scintillator, the maximum recoil energy is 28\% and 22\% of the initial neutron energy, respectively \cite{l2012handbook}.

The liquid scintillator is made of organic fluors dissolved in an organic solvent or aqueous media, with dopant that depends on the application. When gamma interacts with liquid scintillators, the organic solvent or aqueous media absorbs most of the energy due to its higher concentration compared to fluors. After the solvent molecules are excited, they transfer energy to the flour molecules. The fluor molecules then emit photons during their de-excitation. For neutron interaction, the energy transfer to recoil protons or other nuclei will also excite solvent molecules, which then follow the same process of energy transfer and photon emission. The amount of light emitted by a scintillator is proportional to the energy deposit of incident particles \cite{l2012handbook}.

\section{Detector Structure}
\begin{figure} [!ht]
    \centering
    \includegraphics[width=\linewidth]{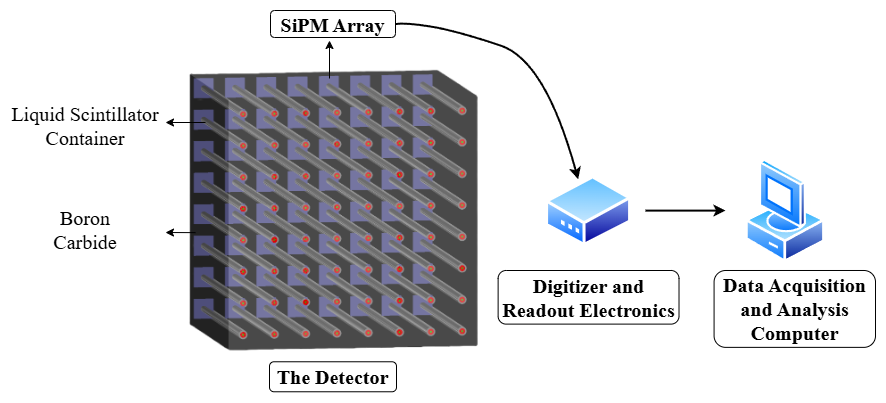}
    \caption{Setup for the directional neutron detector, with SiPM array attache to one end of the detector, a digitizer and read-out electronics, and data acquisition computer.}
    \label{fig:dnd_system}
\end{figure}

This directional neutron detector system is illustrated in Fig.~\ref{fig:dnd_system} and consists of 4 main components: the detector, SiPM array, digitizer and readout electronics, and data acquisition and analysis computer. The detector is composed of a boron carbide moderator, which slows down incoming neutrons and absorbs thermal neutrons, a scintillator that generates photons when energy is deposited, and the SiPM array attached to one end of each scintillator. For the liquid scintillator setup, an optical fiber is placed at the center to guide the scintillation photons to the SiPM array. 

During operation, neutrons enter the detector, some of them directly interact with the scintillator and deposit energy. Others are moderated by the boron carbide, and the resulting thermalized neutrons are captured by boron carbide via the \(\mathrm{^{10}B(n,\:\alpha)^{7}Li}\) reaction, producing 478 keV gammas, which produces 478 keV gammas that can subsequently interact with the scintillator. These two interactions form the basis of the detection mechanism.

The scintillator creates a burst of photons when energy is deposited by gamma or neutron interactions. The optical fiber collects and guides these scintillation photons to the SiPM-coupled end of the scintillator, where they illuminate multiple pixels on the SiPM, where they are converted into electrical signals. Through this approach, the detector can be used for determining the location where particles interact with the detector and thus determine the direction of neutron sources.

\begin{figure}[!ht]
    \centering
    \includegraphics[width=\linewidth]{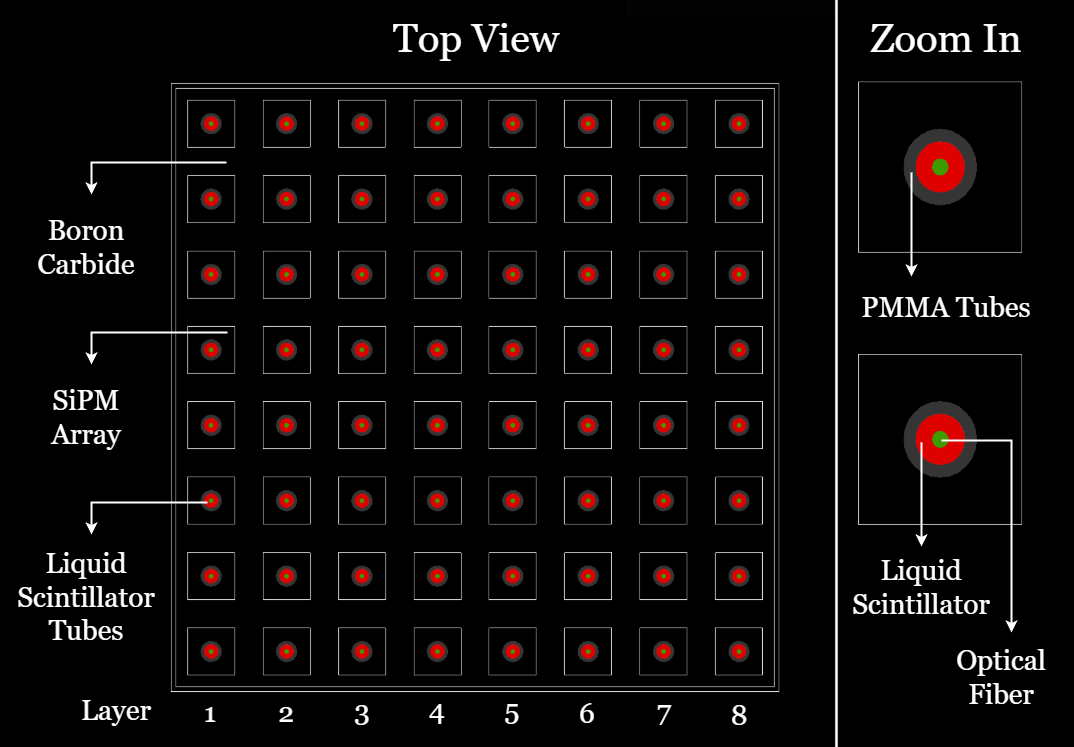}
    \caption{Top view of the detector with liquid scintillators and zoom-in of the liquid scintillator structure in the GEANT4 GUI, showing the geometry and layout of the detector.}
    \label{fig:GEANT4_components}
\end{figure}

The detector geometry includes an aluminum container, boron carbide, scintillator, and optical fibers (in liquid scintillator setup) as shown in Fig.~\ref{fig:GEANT4_components}. The aluminum container is a cube with dimensions of 10~inches on each side. In the setups with liquid scintillators, there are several tubes filled with liquid scintillators inside the container, which we group into different layers. Each of them consists of a PMMA tube with an inner and outer diameter of 6 mm and 9 mm, respectively, and a 2 mm diameter optical fiber placed at the center. The number of liquid scintillators is varied in the simulation to identify the configuration that yields the best detection efficiency and directional sensitivity. A SiPM with a 6 mm by 6 mm detection area, attached to an amplifier board, is placed on one side of each PMMA tube as the photon sensor. Then, the remaining space between the liquid scintillator and the container is filled with boron carbide. In the setup with CsI, the PMMA tubes filled with liquid scintillator are replaced by CsI(Tl) crystals, each shaped with a cross-sectional area of 6 mm by 6 mm and 10 inches long, arranged in the same positions as the PMMA tubes. 
\section{Monte Carlo Simulation}

For the simulation of the directional neutron detector, we use GEANT4 \cite{GEANT4:2002zbu}. The simulation includes the geometry of the detector components and the physics list that defines the physics processes relevant to neutron transport, interactions, and scintillation. The simulation geometry is built based on the actual physical dimensions of the detector components. 

The Shielding physics list in GEANT4 is used to run the simulation, which uses High Precision neutron models for neutron energies below 20 MeV, and includes elastic scattering, inelastic scattering, capture, and fission. The electromagnetic processes are handled by the standard GEANT4 electromagnetic physics, which implement electromagnetic interactions for $\gamma,\,e^{+}/e^{-},\,\mu^{+}/\mu^{-},\,\tau^{+}/\tau^{-}$, and all stable charged hadrons and ions. It is also convenient that the radioactive decay process is implemented. Note that this physics list doesn’t include optical photon transportation and scintillation process, so we added the \textit{G4OpticalPhysics} module into the simulation.

\textit{G4GeneralParticleSource} is used to generate neutron particles with specific energies and directions. The source is placed outside the detector, 50 cm away from its center, resulting in a 37.3 cm distance from the source to the detector surface. We set the angular distribution to be isotropic and limit the solid angle to save simulation time but large enough to fully cover the detector. We also set an energy threshold of 1 keV, so any energy deposits smaller than 1 keV will be discarded.

\section{Material and Geometry Optimization}

In this section, we discuss the effects of different materials and setups on detector performance, aiming to identify the optimal detector configuration for directional neutron detection.

\subsection{Evaluation of Detector Materials}

Although this detector design is intended to use boron carbide as a neutron moderator and liquid scintillator as the scintillator, identifying the optimal material requires comparing its effectiveness with other candidate materials.

We evaluated three material configurations to compare detector performance: (1) boron carbide as the moderator with liquid scintillator, (2) boron carbide with a CsI scintillator, known for its high gamma sensitivity, and (3) liquid scintillator only, without a moderator. The third setup offers potential benefits in reduced weight and cost, and helps assess the necessity of boron carbide for efficient neutron detection.

To maximize the detector performance, we systematically studied the effects of different materials using GEANT4 simulations. First, we evaluated the detection efficiency for three different setups using up to 2 MeV monoenergetic neutron source with a standard 8 $\times$ 8 array of liquid scintillators evenly distributed in the container. The detection efficiency is defined as:

\begin{equation}
    efficiency\: = ~ \frac{N_{detected}}{N_{total}}
\end{equation}

Where \(\mathrm{N_{detected}}\) is the number of neutrons that deposited more than 1 keV in the scintillator and \(\mathrm{N_{total}}\) is the total emitted neutrons.

Tab.~\ref{tab:efficiency} shows the detection efficiency for three different setups. Boron carbide with liquid scintillator gives great efficiency across most of the fast neutron energy range. For neutron energy smaller than 1 MeV, the setup with only liquid scintillator gives better detection efficiency. However, it decays much faster than other setups at higher neutron energies. This is most likely due to the lack of ability to slow down neutrons.

\begin{table}[h!]
\centering
\renewcommand{\arraystretch}{0.9} 
\begin{tabular}{@{\hskip 1pt}c|@{\hskip 1pt}c@{\hskip 1pt}c@{\hskip 1pt}c@{\hskip 1pt}}
\hline
\begin{tabular}[c]{@{}c@{}}Neutron\\ Energy (MeV)\end{tabular} &
  \begin{tabular}[c]{@{}c@{}}\\$B_4C$ + $LS$\end{tabular} &
  \begin{tabular}[c]{@{}c@{}}Setup Efficiency (\%)\\$B_4C$ + $CsI$\end{tabular} &
  \begin{tabular}[c]{@{}c@{}}\\$LS$ \end{tabular} \\ \hline
\(\lesssim 0.5\) & 17.72 & 15.89 & 22.23 \\
\(\lesssim 1.0\) & 18.26 & 15.02 & 17.03 \\
\(\lesssim 2.0\) & 14.95 & 12.59 & 12.37 \\
\(\lesssim 3.0\) & 13.09 & 11.51 & 9.96 \\ \hline
\end{tabular}
\caption{Detection efficiency for the three different setups at selected neutron energies.}
\label{tab:efficiency}
\end{table}

\begin{figure} [!ht]
    \centering
    \includegraphics[width=1\linewidth]{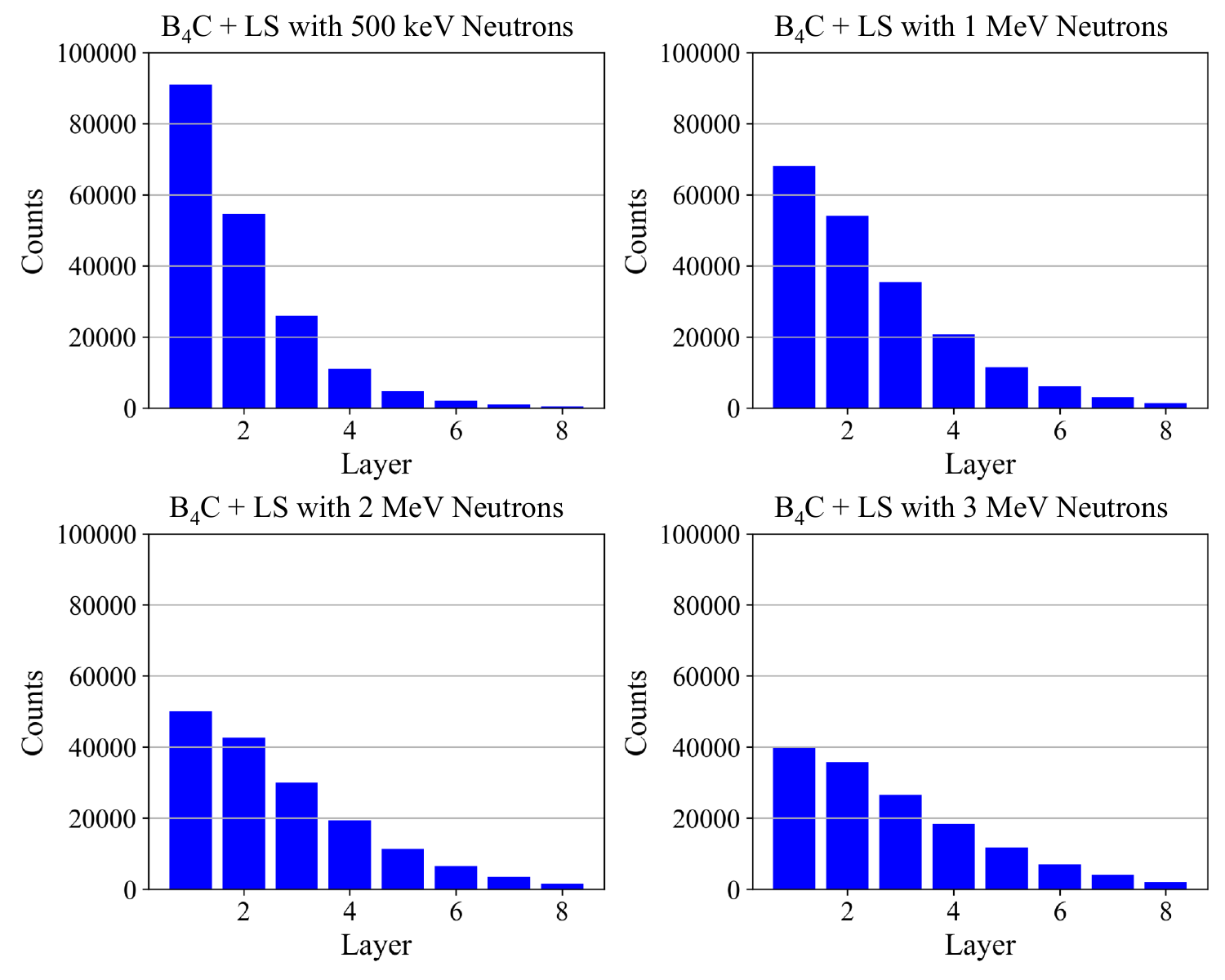}
    \caption{Number of events detected per layer, for the setup of liquid scintillator combine with boron carbide, for several selected neutron energies in the fast neutron range.}
    \label{fig:b4c_ls}
\end{figure}

\begin{figure} [!ht]
    \centering
    \includegraphics[width=1\linewidth]{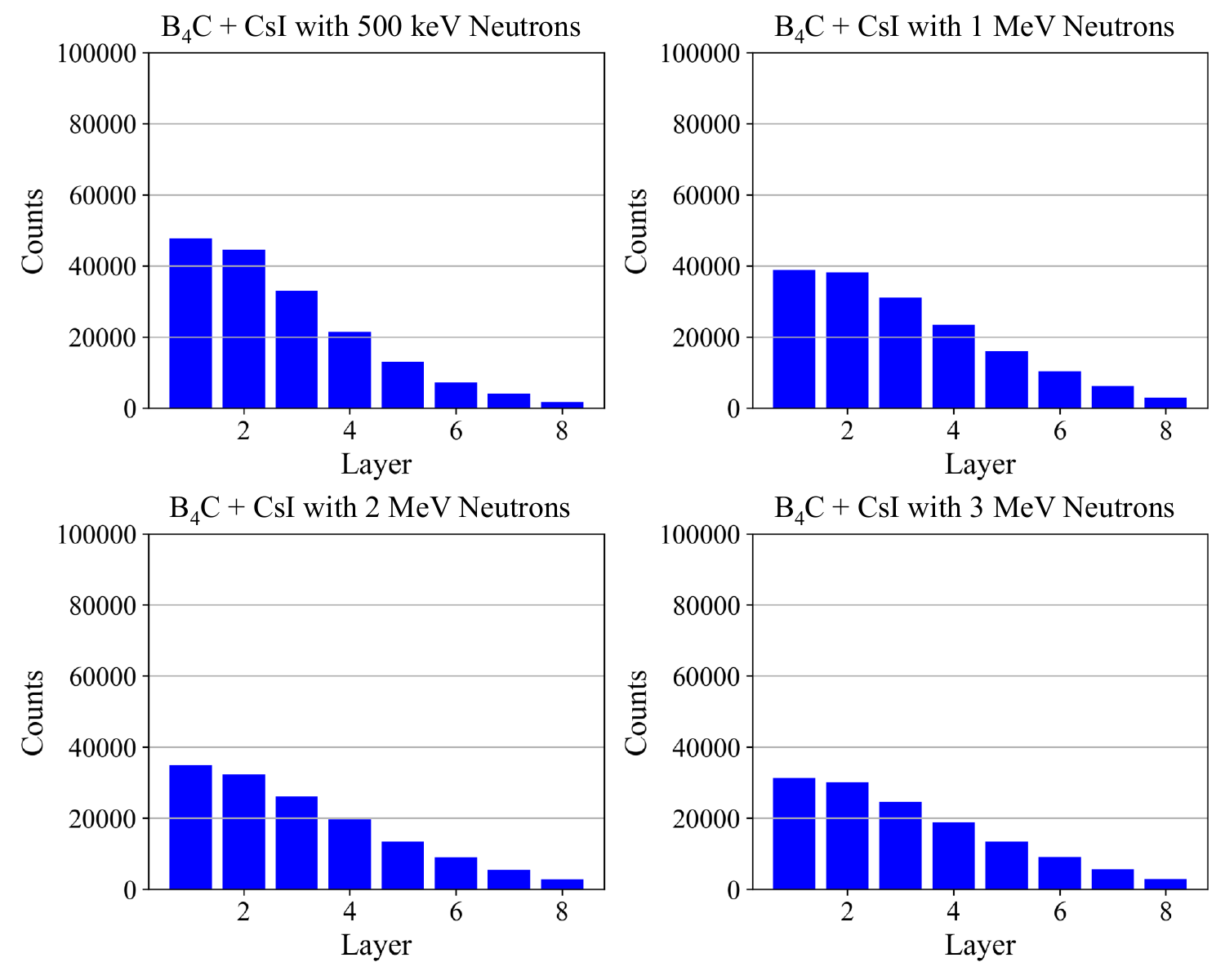}
    \caption{Number of events detected per layer, for the setup of CsI combine with boron carbide, for several selected neutron energies in the fast neutron range.}
    \label{fig:b4c_csi}
\end{figure}

\begin{figure} [!ht]
    \centering
    \includegraphics[width=1\linewidth]{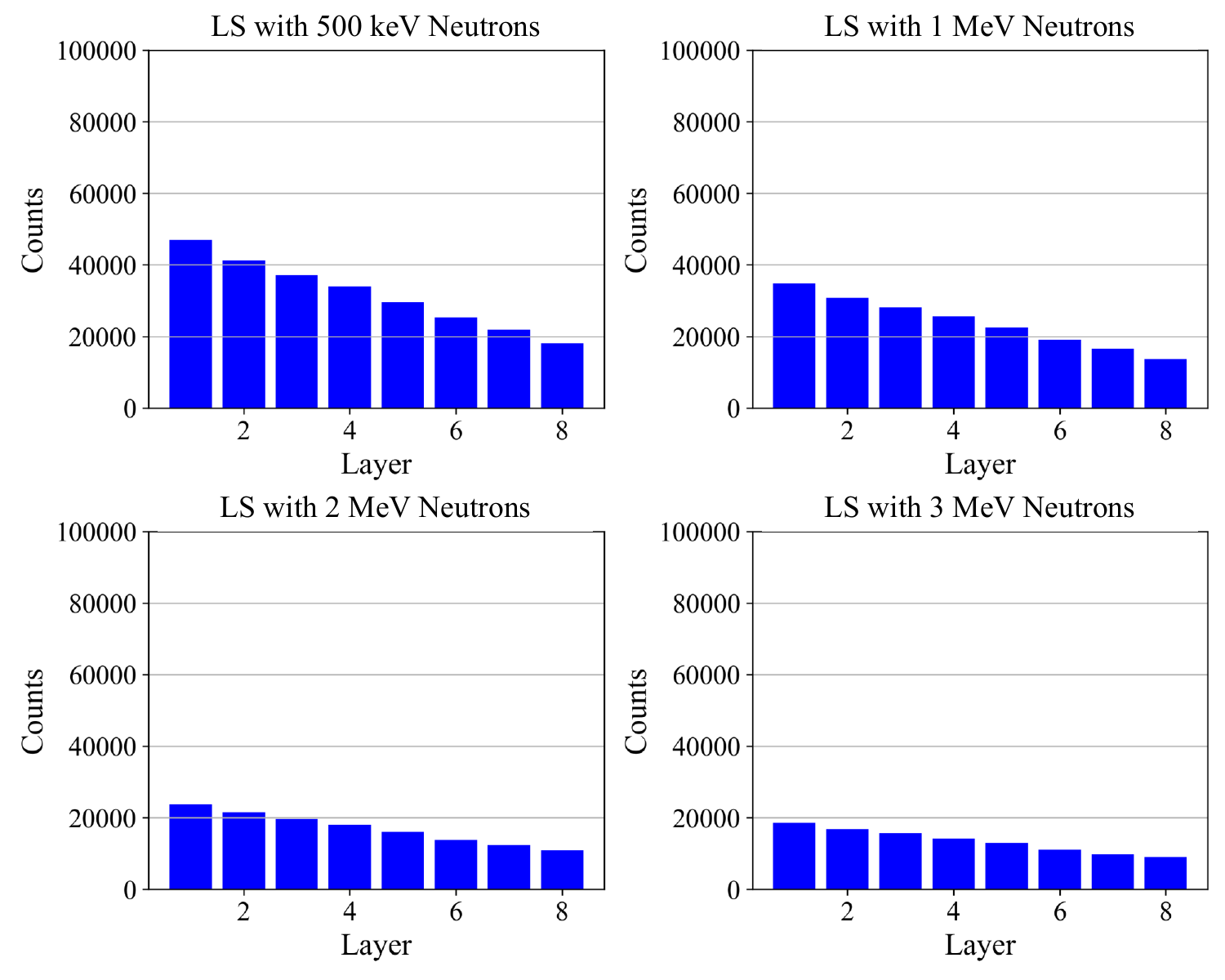}
    \caption{Number of events detected per layer, for the setup of only liquid scintillator, for several selected neutron energies in the fast neutron range.}
    \label{fig:onlyls}
\end{figure}

In addition to detection efficiency, the ability to determine the neutron direction is also important for the detector design. For this purpose, we evaluate the direction sensitivity of the same three detector setups. To quantify the directional sensitivity, we compare the number of energy deposition events in each layer from the layer facing the neutron source to the layer on the opposite side of the detector. If the difference between layers is large, it indicates that the detector has a good ability to recognize the neutron direction. Fig.~\ref{fig:b4c_ls} to~\ref{fig:onlyls} show the comparison of the number of energy deposition events in different layers. We can then fit the distribution, which gives us an idea of the detector's directional sensitivity. Again, for the simulation, we use several different neutron energies in the fast neutron range, with an 8 $\times$ 8 array of liquid scintillator containers.

\begin{figure} [!ht]
    \centering
    \includegraphics[width=\linewidth]{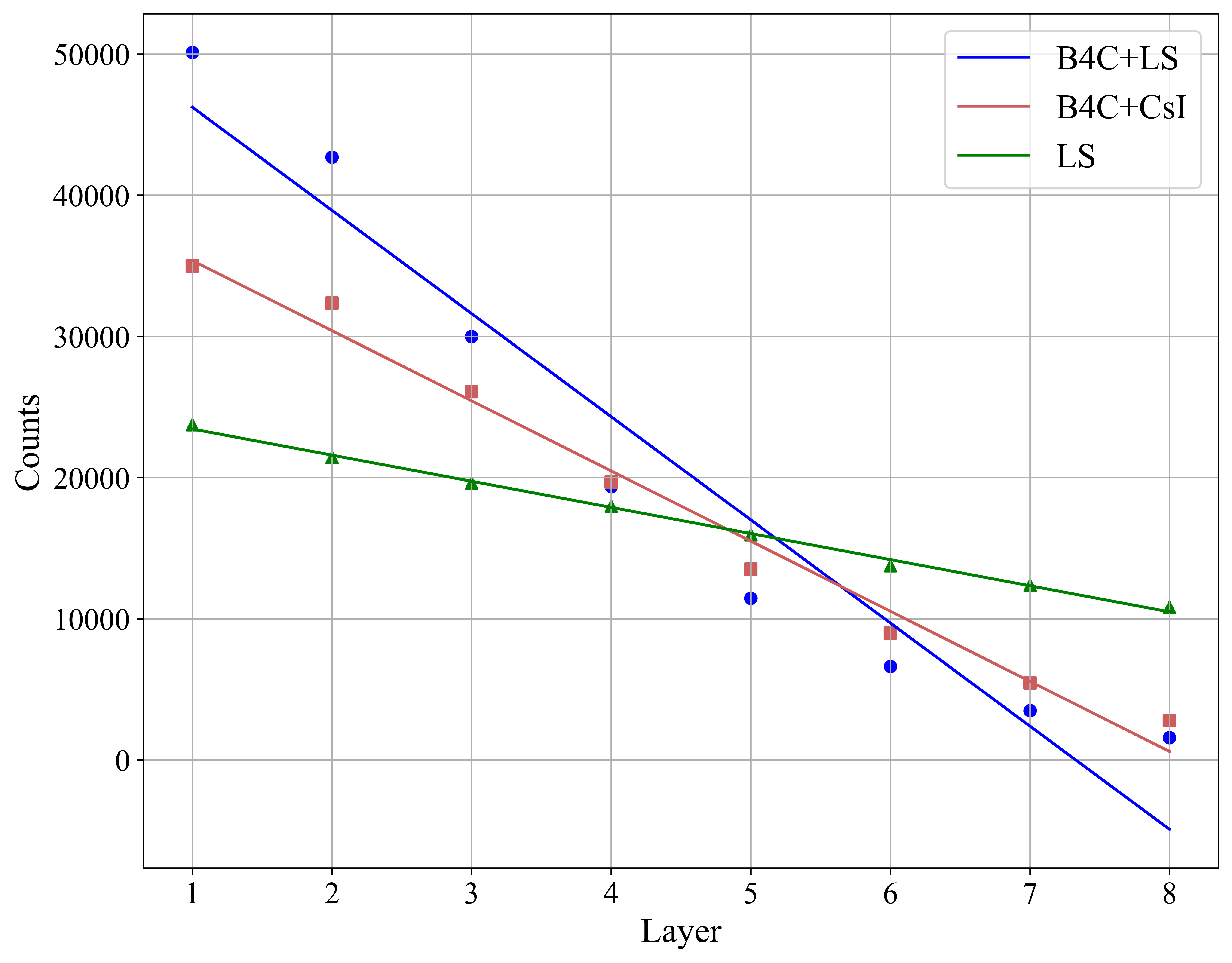}
    \caption{Fitting the data for 2 MeV incident neutrons with different setups to obtain their spatial variation gradient $dN/dx$.}
    \label{fig:fom_slope}
\end{figure}

Fig.~\ref{fig:fom_slope} shows the fitted results for the 2 MeV neutron source in different setups, along with the rate $dN/dx$, which serves as a proxy for directional sensitivity. A steeper gradient suggests better sensitivity to the incident direction. Tab.~\ref{tab:dndx} shows the $dN/dx$ for all the different neutron energies and setups.

\begin{table}[]
\centering
\begin{tabular}{@{\hskip 1pt}c|@{\hskip 4pt}c@{\hskip 4pt}c@{\hskip 4pt}c@{\hskip 1pt}}
\hline
\begin{tabular}[c]{@{}c@{}}Neutron\\ Energy (MeV)\end{tabular} &
  \begin{tabular}[c|]{@{}c@{}}\\ $B_4C$ + LS\end{tabular} &
  \begin{tabular}[c]{@{}c@{}}Setup \(|dN/dx|\)\\$B_4C$ + $CsI$\end{tabular} &
  \begin{tabular}[c|]{@{}c@{}}\\ $LS$ \end{tabular} \\ \hline
0.5 & 0.012(3) & 0.0073(7) & 0.0040(1) \\
1.0 & 0.0098(1) & 0.0057(3) & 0.0030(1) \\
2.0 & 0.0073(1) & 0.0050(3) & 0.0019(1) \\
3.0 & 0.0059(1) & 0.0045(2) & 0.0014(1) \\ \hline
\end{tabular}
\caption{Normalized spatial gradients \( |dN/dx| \) for three detector configurations at various incident neutron energies. Each value is normalized per incident neutron, and the statistical uncertainties are derived from simulations with \( 10^6 \) events. Numbers in parentheses indicate the uncertainty in the last digits.}
\label{tab:dndx}
\end{table}

\subsection{Evaluation of Layer Configuration}

To maximize the neutron detection efficiency of the proposed directional detector, we study the effect of the number of scintillators per layer using Monte Carlo simulations. First, we vary the number of liquid scintillators from 8 to 16 while keeping the detector volume and other parameters constant. In this study, we use boron carbide as the moderator and liquid scintillator as the scintillating material, with a 2 MeV neutron source.

\begin{figure} [!ht]
    \centering
    \includegraphics[width=0.96\linewidth]{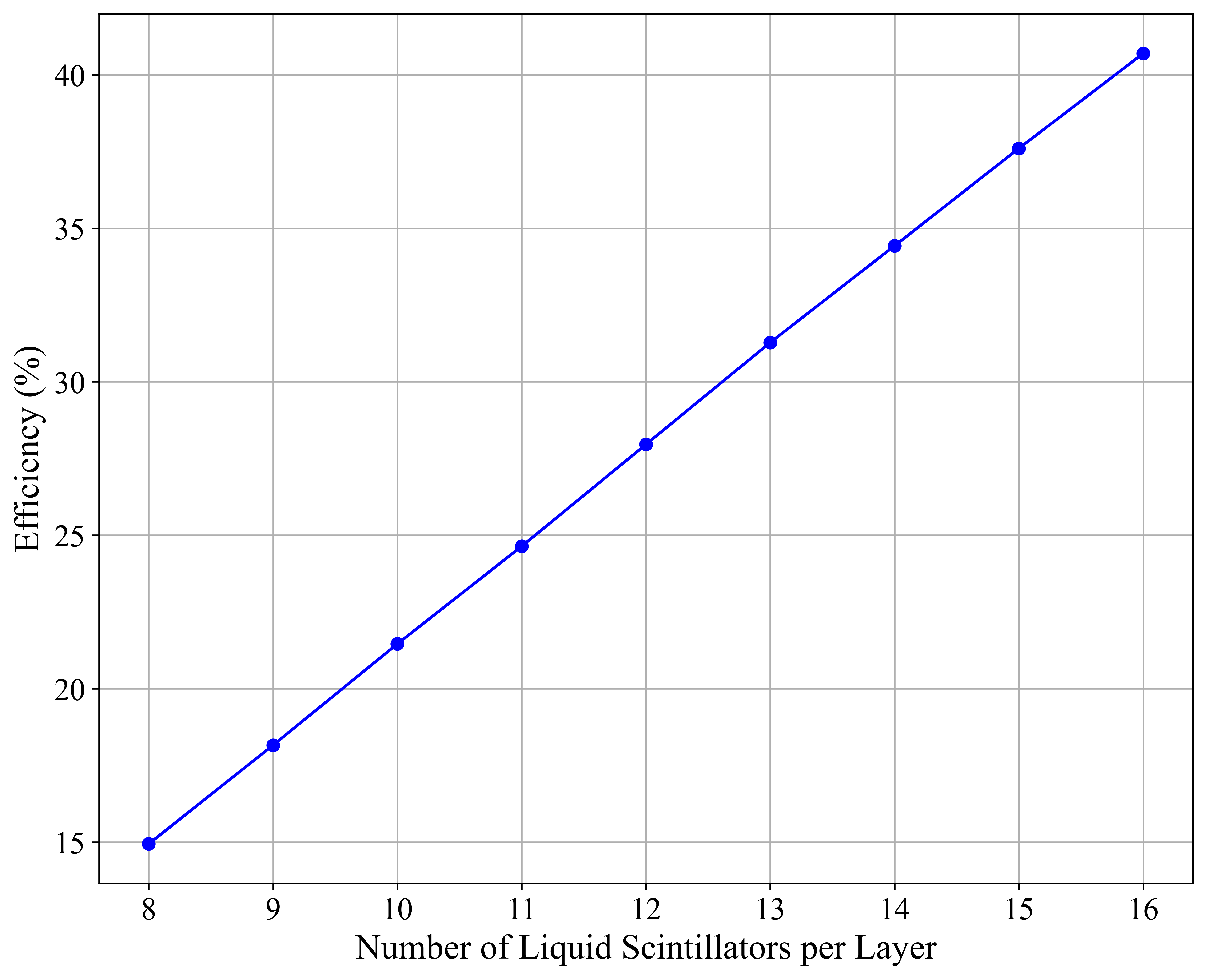}
    \caption{Detection efficiency with different number of liquid scintillators per layer, evaluated with 2 MeV incident neutrons.}
    \label{fig:diff_tube_num}
\end{figure}

\begin{figure} [!ht]
    \centering
    \includegraphics[width=\linewidth]{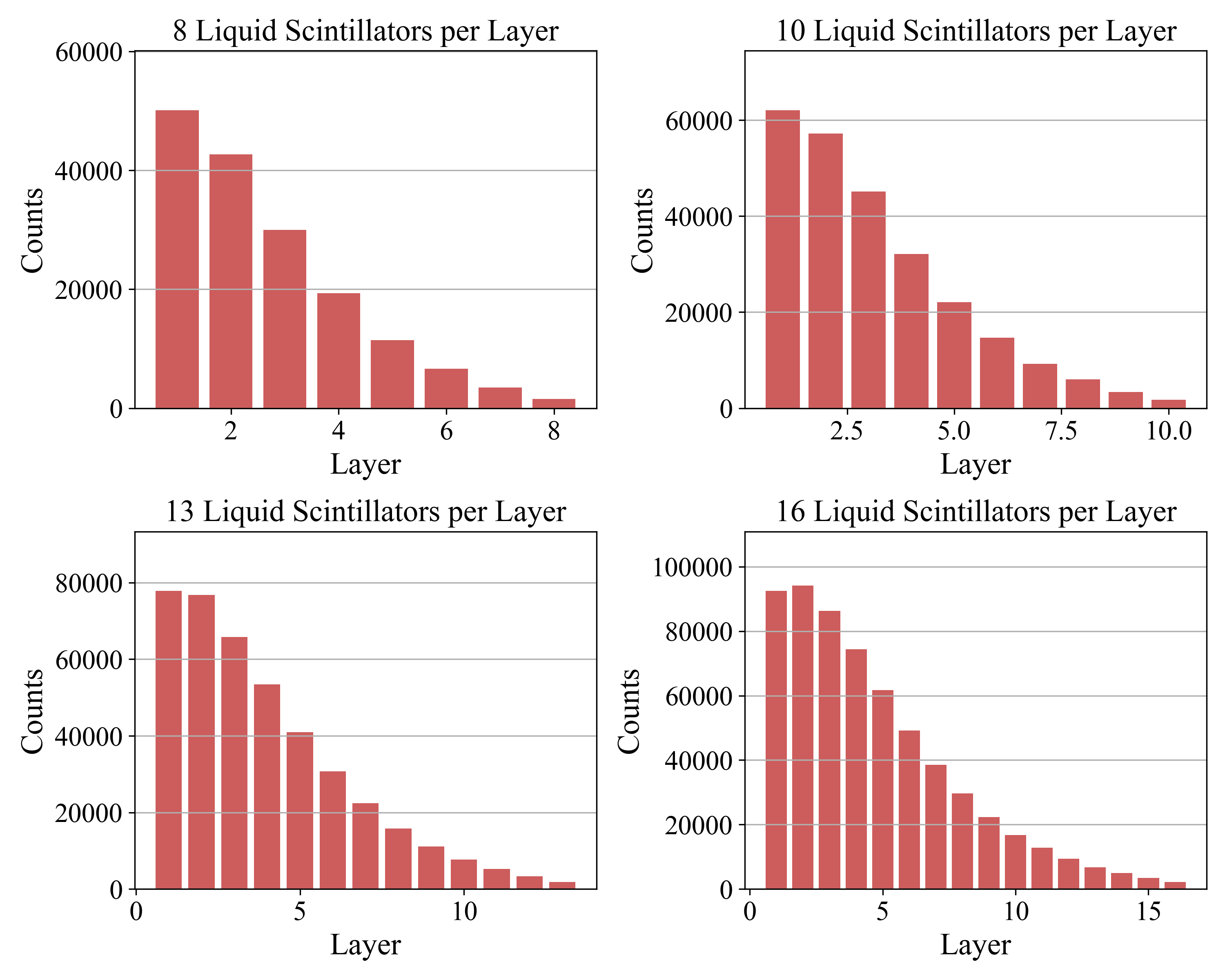}
    \caption{Hit counts per layer for different numbers of liquid scintillators. With increasing number of liquid scintillators, the counts in the second layer are larger and they eventually exceed those in the first layer, evaluated with 2 MeV incident neutrons.}
    \label{fig:diff_tube_num_counts}
\end{figure}

Fig.~\ref{fig:diff_tube_num} illustrates that the detector's efficiency increases nearly linearly with the number of scintillators per layer, suggesting a general trend toward improved performance with higher granularity. However, factors such as cost, readout complexity, optical cross-talk, and directional resolution must also be considered. As shown in Fig.~\ref{fig:diff_tube_num_counts}, increasing the number of scintillators can shift the peak response to the second layer due to spatial rearrangement, potentially degrading directional sensitivity. Balancing these considerations, a configuration with approximately 11 scintillators per layer with about 25\% efficiency for 2 MeV neutrons appears to offer an optimal trade-off between efficiency and directional accuracy.

\section{Evaluation of Neutron Source Directionality Detection}

As it explained in previous sections, to assess the directional sensitivity of the proposed detector, we conducted a series of GEANT4-based Monte Carlo simulations in which neutron sources were placed at known locations relative to the detector array. The spatial distribution of energy depositions within the liquid scintillator was recorded for each incident neutron event, capturing the interaction topology essential for inferring directionality. 

\begin{figure} [!ht]
    \centering
    \includegraphics[width=\linewidth]{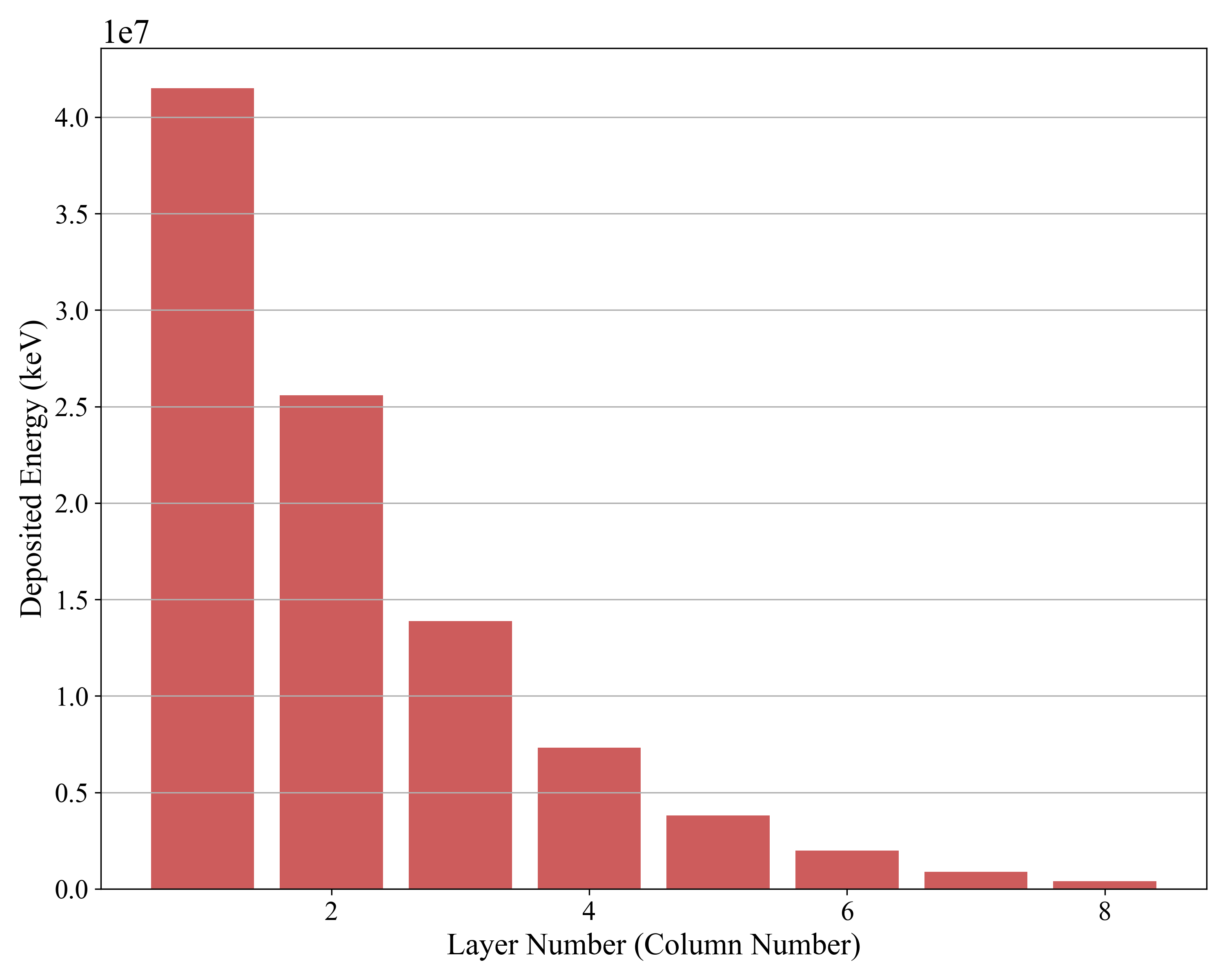}
    \caption{Deposited energy distribution was analyzed across all layers, counted from the side nearest to the neutron source to the farthest, for neutrons with an energy of 2~MeV and a total of 2 million simulated events.}
    \label{fig:Energy Distribution}
\end{figure}

The directional response of the detector was evaluated by analyzing the asymmetry in signal intensity across the SiPM-coupled fiber array. Specifically, the total energy deposited in each layer was computed. For clarity, we define a layer as all liquid scintillator located at the same coordinate along one spatial dimension (e.g., all liquid scintillators sharing the same \(x\)-position). For the coordinate system used in this study, we defined the \(x\) and \(y\) axes as perpendicular to the orientation of the scintillator, and the \(z\) axis as parallel to the tube length. To simplify the analysis, our evaluation of directional sensitivity focused on neutron sources incident along the \(x\) and \(y\) directions. Extension of this analysis to include the \(z\)-axis will be addressed in future work. By projecting the energy deposition profiles along the \(x\), and \(y\), we established directional vectors that indicate the likely origin of the neutron flux. When a neutron source was positioned adjacent to a particular face of the detector cube (e.g., the ``right'' or ``top'' side), a statistically significant increase in energy deposition was consistently observed in the layers nearest to the source. This spatial gradient in energy deposition served as a reliable and reproducible signature for determining the direction of the incoming neutron radiation (Fig.~\ref{fig:Energy Distribution}). 

To further quantify the angular resolution, we trained a supervised machine learning classifier using the spatial gradient in energy, hit distribution patterns, and energy deposition in layers as input features. 
Let \(\mathbf{x}_i = \left[g_i, h_i, e_i\right]\) be the feature vector for the \(i\)-th event, where:
\begin{itemize}
    \item \(g_i\) is the spatial gradient of deposited energy,
    \item \(h_i\) is the hit distribution pattern,
    \item \(e_i\) is the vector of energy depositions across layers.
\end{itemize}

We trained a supervised binary classifier \(f(\mathbf{x}_i; \boldsymbol{\beta})\) based on logistic regression to predict whether a neutron source originates from a particular direction. The model outputs the probability that the class label \(y_i \in \{0, 1\}\) is equal to 1, given the input features \(\mathbf{x}_i\), as follows:
\[
P(y_i = 1 \mid \mathbf{x}_i; \boldsymbol{\beta}) = \sigma(\boldsymbol{\beta}^\top \mathbf{x}_i) = \frac{1}{1 + \exp(-\boldsymbol{\beta}^\top \mathbf{x}_i)},
\]
Where \(\sigma(\cdot)\) is the sigmoid activation function and \(\boldsymbol{\beta}\) denotes the learnable parameters of the model.

The predicted label is obtained by thresholding the output probability:
\[
\hat{y}_i = 
\begin{cases}
1 & \text{if } P(y_i = 1 \mid \mathbf{x}_i; \boldsymbol{\beta}) \geq 0.5, \\
0 & \text{otherwise}.
\end{cases}
\]

The model is trained by minimizing the binary cross-entropy loss:
\begin{align}
\mathcal{L}(\boldsymbol{\beta}) = -\sum_{i=1}^N \bigg[ 
    & y_i \log P(y_i = 1 \mid \mathbf{x}_i; \boldsymbol{\beta}) \notag \\
    & + (1 - y_i) \log \left(1 - P(y_i = 1 \mid \mathbf{x}_i; \boldsymbol{\beta}) \right) 
\bigg].
\end{align}

\begin{figure} [!ht]
    \centering
    \includegraphics[width=\linewidth]{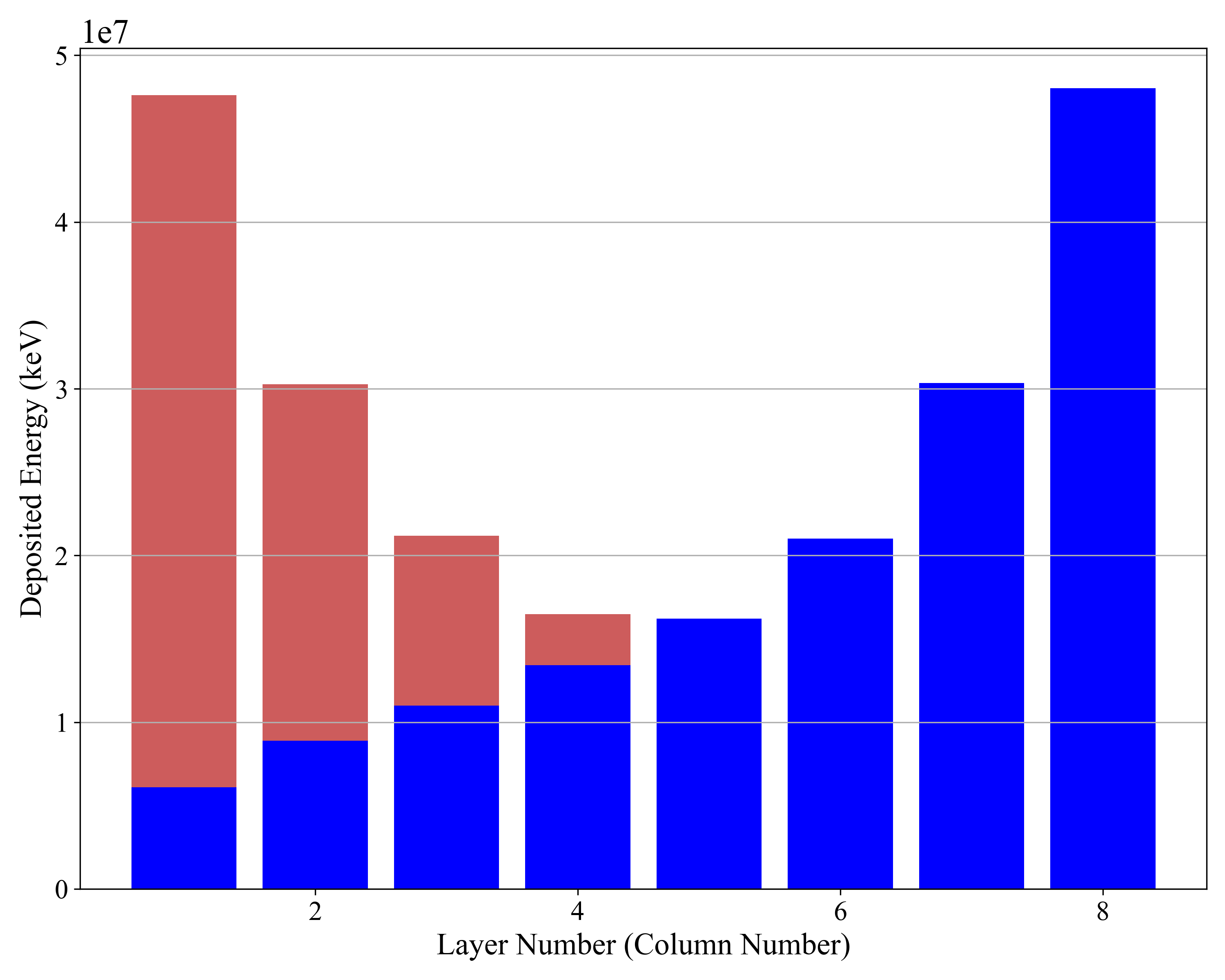}
    \caption{Deposited energy distribution across all layers and columns, ordered from the side nearest to the neutron source to the farthest. The blue bars represent energy deposited in layers aligned along the \(x\)-axis, with the first layer (Layer 1) located on the \(-x\) side receiving the highest energy. The red bars correspond to columns along the \(y\)-axis, where Column 8 (located on the \(+y\) side) shows the highest energy deposition among all columns. This asymmetry reflects the spatial distribution of neutron interactions relative to the source direction.}

    \label{fig:Energy Distribution2}
\end{figure}

The logistic regression model achieved 100\% classification accuracy in identifying the neutron source direction among the four cardinal axes (\(\pm x\), \(\pm y\)). Its performance remained stable across varying neutron energies and moderate background conditions, supporting the feasibility of the proposed detection method for practical use. Initially trained on sources facing a single detector face, the model was later generalized by expanding the training dataset to include flux directions approaching the detector edges and corners. As shown in Fig.~\ref{fig:Energy Distribution2}, the energy deposition pattern from a neutron flux directed toward the \(-x\) and \(+y\) edge demonstrates the model’s capacity to resolve non-orthogonal incident angles. The feature vector prioritizes the spatial gradient (\(g\)) with the highest learned weight (\(\beta\)), followed by hit distribution (\(h\)). Classification proceeds in two steps: first, by narrowing down possible directions based on feature evaluation, and then by applying the trained logistic regression model to select the most probable direction. This approach enables accurate source localization even in complex geometries.

These results validate the proposed detector’s capability to discern neutron source directionality with high accuracy using compact, portable hardware. This approach offers a promising solution for field-deployable systems in security scanning, nuclear nonproliferation, and environmental monitoring scenarios.

\section{Conclusion}

This work presented a simulation-based study of a compact directional neutron detector using boron carbide moderation, scintillators, and SiPM readout. GEANT4 simulations were used to optimize material configuration and detector geometry, finding neutron detection efficiencies ranging from approximately 10\% to 30\%.

Directional sensitivity was demonstrated by analyzing energy deposition patterns and training a machine learning model on spatial features. The classifier achieved 100\% accuracy in identifying source direction among four cardinal axes and remained robust under varying neutron energies and complex source positions.

These results confirm the feasibility of the proposed design, which combines liquid scintillator with boron carbide, for practical applications in security, environmental monitoring, and scientific research. Additionally, the CsI-based alternative also showed encouraging results and may be worth further investigation. Future work will focus on extending directionality analysis and hardware prototyping.

\section{Acknowledgements}
The work presented in this paper (NOT THESIS) has been supported by funding from the Mitchell/Heep Endowment for Experimental High Energy Physics, and by LANL - TRIAD funds provided for Los Alamos A\&M Dark Matter and Neutrino Alliance (LAAMDNA). We also acknowledge funding support from the Mitchell Institute.

\bibliographystyle{elsarticle-num}
\bibliography{dnd_sim}


\end{document}